\documentclass[cameraready]{Interspeech}
\usepackage{amsmath}
\usepackage{multirow}
\usepackage{graphicx}
\usepackage{bm}
\usepackage{algorithm}
\usepackage{algorithmic}
\usepackage{amsmath}
\usepackage{booktabs}
\usepackage{graphicx}
\usepackage{hyperref}
\usepackage{wasysym}
\usepackage{multirow}
\usepackage{booktabs}
\usepackage{pifont}
\usepackage{setspace}
\usepackage{subcaption}
\usepackage{indentfirst}
\newcommand{\myparagraph}[1]{\vspace{6pt}\noindent\textbf{#1.}\quad}

\title{Domain-Adaptive Dual-Gating Mixture of Experts for Generalizable Speech Deepfake Detection}

\author[affiliation={1}]{Siqing}{Qin}
\author[affiliation={2}]{Zhe}{Li}
\author[affiliation={1}, correspondingauthor]{Kong Aik}{Lee}
\author[affiliation={1}]{Man-Wai}{Mak}

\address{
  $^1$ Dept. of Electrical and Electronic Engineering, The Hong Kong Polytechnic University \\
  $^2$ Speech, Language, and Cognition Laboratory, The University of Hong Kong
}

\email{siqing.qin@connect.polyu.hk}


\usepackage{comment}

\begin{document}

\maketitle
\nolinenumbers
\begin{abstract}
  Recent advances in speech deepfake detection (SDD) have leveraged the Mixture of Experts (MoE) to enhance generalization capacity. However, existing gating networks often overlook the acoustic and temporal cues of deepfakes. In this work, we propose a novel domain-adaptive dual-gating MoE (DADGMoE) framework for SDD under unseen attack types and acoustic conditions. Our innovative dual-gating mechanism leverages Sinc-layer-based filters to process both low-level acoustic signals (raw waveforms) and high-level speech representations from a large self-supervised learning (SSL) model. It further incorporates domain prototypes to guide expert routing based on implicit deepfake patterns. The lightweight affine experts process the routed inputs. Experiments show that our DADGMoE significantly outperforms the baseline, achieving up to a 40.8\% relative EER reduction on challenging out-of-dataset benchmarks. This framework demonstrates superior generalization capabilities and efficient design.
\end{abstract}

\section{Introduction}

Speech deepfake detection (SDD) aims to counter the growing threats posed by realistic deepfakes~\cite{asvspoof2019, asvspoof5, add2022, yi2023add}. While these systems achieve remarkable performance under controlled conditions~\cite{aasist2024, truong2025addressing}, their generalization capabilities against unseen attack types, varied acoustic environments, or novel deepfake codecs remain a critical challenge~\cite{crs_ds3, huang25sharp}. This challenge stems from the unique characteristics of diverse deepfake generation methods and multilingual data, which essentially form distinct domains. Consequently, developing flexible detectors capable of handling such increasingly variable scenarios is crucial.

To tackle this challenge, Mixture of Experts (MoE) has emerged as a promising method in SDD~\cite{moe2025icassp, pan2025moe, wang2025moe, laakkonen2025moe, hao2025moeijcnn}. An MoE is a hierarchical model comprising an ensemble of $N$ experts, where a probabilistic gating function $\mathcal{G}$ dynamically determines each expert's contribution to the final output~\cite{moe2025survey,li2026towards}. While conceptually well-suited for handling diverse input characteristics and improving generalization, the gating mechanisms in existing MoE for audio tasks predominantly rely on generic feed-forward networks (FFNs)~\cite{moe2025icassp, hao2025moeijcnn}. Such generic gates often overlook the unique acoustic and temporal artifacts embedded within deepfake audio. These artifacts may range from subtle physical distortions to high-level prosodic inconsistencies. We argue that a more effective gating function must leverage speech-specific artifacts to direct distinct deepfake patterns to specialized experts, thereby enhancing the model's ability to generalize across diverse, unseen conditions.

\begin{figure}
  \centering
  \includegraphics[width=0.9\linewidth]{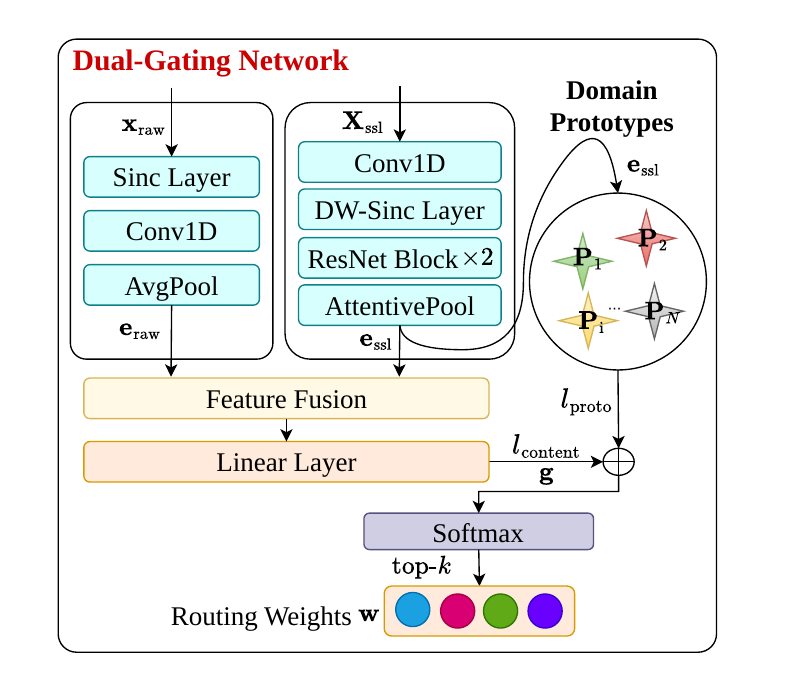}
  \caption{Illustration of the proposed dual-gating network, highlighting the novel Sinc layer-based artifact filtering on both raw waveforms and SSL features, and the end-to-end learned domain prototypes for routing.}
  \label{fig:dadg}
\end{figure}

To this end, we propose a novel domain-adaptive dual-gating Mixture of Experts (DADGMoE) framework. Central to our approach is an innovative dual-gating mechanism (Figure~\ref{fig:dadg}) that comprehensively processes both raw waveforms and high-level self-supervised learning (SSL) features commonly used in SDD. For speech-specific artifact filtering, we incorporate a variant of the Sinc layer into our dual-gating mechanism. Sinc layers have demonstrated exceptional capabilities in modeling data-driven band-pass filters directly from raw audio, making them ideal for capturing subtle low-level physical artifacts~\cite{sincnet, sincnet2, rawnet2}. Crucially, we extend their applications by employing a depthwise SincNet on SSL feature sequences to dynamically filter high-level temporal dynamics.
This dual application of SincNet enables our gate to comprehensively perceive deepfake artifacts across SSL features and spectra.

\begin{figure}[t]
  \centering
  \includegraphics[width=\linewidth]{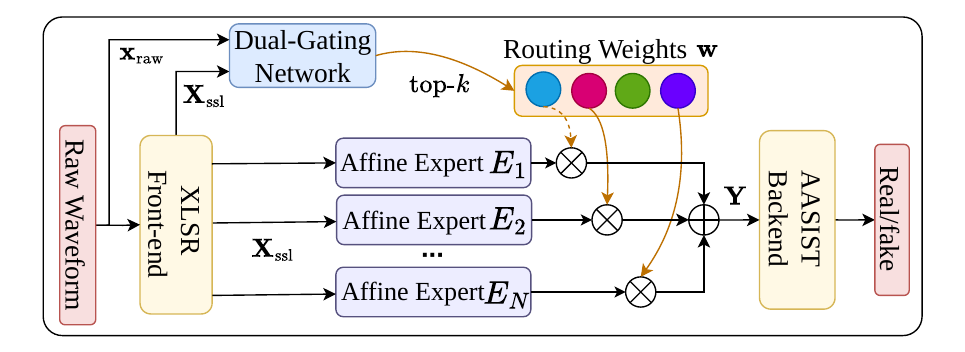}
  \caption{Overview of the proposed DADGMoE framework.}
  \label{fig:overview}
\end{figure}

Furthermore, our DADGMoE framework employs a set of highly lightweight affine experts, each comprising a single Batch Normalization (BN) layer, as shown in Figure~\ref{fig:overview}. This design is inspired by the understanding that BN layers, through their learnable affine parameters ($\bm{\gamma}$ and $\bm{\beta}$), inherently encode domain-specific information~\cite{chang2019bn, wu2024testbn}. Thus, each affine expert can efficiently adapt to the unique statistical distributions of specific deepfake patterns or domains routed by the gating network.
Moreover, to explicitly address the challenge of diverse attack types without relying on the seen domain labels, we introduce learnable domain prototypes within the gating mechanism. These prototypes automatically discover and anchor distinct deepfake patterns in the feature space, guiding the router to assign inputs to the most statistically compatible affine experts.

Experimental results demonstrate the superiority of our proposed framework. Evaluated on the ASVspoof 2021 Deepfake (21DF)~\cite{yamagishi2021asvspoof}, In-the-Wild (ITW)~\cite{muller2022itw}, and Fake-or-Real (FoR)~\cite{reimao2019for} datasets, our method significantly outperforms the strong XLSR-AASIST baseline. Specifically, the optimal configuration using dual-gating with top-$k$ routing ($k=2$) achieves relative equal error rate (EER) reductions of 61.7\% on the 21DF set, 43.3\% on the challenging ITW set, and 40.8\% on the FoR set.
These substantial improvements
validate the effectiveness of our dual-gating mechanism and domain prototype-guided experts in enhancing domain generalization.

\section{Proposed Method}
Figure~\ref{fig:overview} illustrates the proposed \textbf{DADGMoE} framework for generalizable SDD. Given an input utterance, the method outputs a bonafide/spoof prediction under unseen attack types and acoustic conditions. DADGMoE appends an MoE module with $N$ lightweight affine experts $\mathcal{E}=\{E_1, E_2,\dots, E_N\}$ to an SSL-based backbone, where $N$ denotes the number of experts. The module computes sample-dependent routing weights and applies the selected experts to transform the intermediate SSL feature.
\vspace{-0.1cm}
\subsection{Sinc-Based Acoustic Filtering}
\vspace{-0.1cm}
\label{subsec:preliminaries}
Standard convolutional neural networks (CNNs) process raw audio using learnable convolutional filters of kernel length $K$. Because these filters are learned in an unconstrained manner, they are not explicitly parameterized by frequency characteristics, making them difficult to interpret as band-pass filters.

SincNet~\cite{sincnet} addresses this limitation by constraining the first-layer kernels to represent learnable band-pass filters. The $k$-th Sinc filter is parameterized by two scalar cutoff frequencies, a lower cutoff $f_{1,k}$ and an upper cutoff $f_{2,k}$. These parameters deterministically specify a band-pass impulse response (i.e., a discrete-time convolution kernel) $\bm{h}_k(f_{1,k},f_{2,k})$, whose coefficients are computed from $(f_{1,k},f_{2,k})$ and then multiplied by a Hamming window. Given an input signal $\bm{x}_{\text{raw}}$, the filter output $\bm{y}_k$ is computed by 1D convolution:
\begin{equation}
  \bm{y}_k = \bm{x}_{\text{raw}} \circledast \bm{h}_k(f_{1,k}, f_{2,k}),
\end{equation}
where $\circledast$ denotes 1D convolution. This parameterization reduces the number of learnable parameters per filter from $K$ to 2 and provides a direct mapping between each filter and a frequency band. In our framework, we adopt this formulation to extract low-level artifact cues from raw waveforms and extend it to high-level SSL features via a depthwise Sinc layer, as described below.
\vspace{-0.1cm}
\subsection{Domain Adaptive Dual Gating Mechanism}
\vspace{-0.1cm}
\label{sec:dual_gating}
The gating network (Figure~\ref{fig:dadg}) outputs routing weights over experts. It consists of two parallel branches that extract gating embeddings from the same input utterance: a waveform-based branch that targets low-level physical artifacts and an SSL-feature-based branch that targets higher-level inconsistencies.
\vspace{-0.1cm}
\subsubsection{Waveform-Based Gating}
\vspace{-0.1cm}
This branch processes the raw waveform $\bm{x}_{\text{raw}} \in \mathbb{R}^{T}$, where $T$ denotes the number of samples in the input signal. It first applies a Sinc layer~\cite{sincnet} whose kernels implement learnable band-pass filters parameterized by cutoff frequencies. We denote the resulting waveform Sinc filter bank by $\mathcal{H}_{\text{sinc}}^{\text{raw}}=\{\bm{h}_k(f_{1,k},f_{2,k})\}_{k=1}^{N_{\text{raw\_filters}}}$. The output feature map is computed as
\begin{equation}
  \bm{F}_{\text{sinc}}^{\text{raw}} =
  \bm{x}_{\text{raw}} \circledast \mathcal{H}_{\text{sinc}}^{\text{raw}}.
\end{equation}
Then, a Conv1D layer followed by global average pooling yields a fixed-dimensional embedding:
\begin{equation}
  \bm{e}_{\text{raw}} =
  \mathrm{AvgPool}\left(
    \mathrm{Conv1D}(\bm{F}_{\text{sinc}}^{\text{raw}})
  \right) \in \mathbb{R}^{D},
\end{equation}
where $\mathrm{Conv1D}(\cdot)$ denotes a learnable 1D convolutional layer, $\mathrm{AvgPool}(\cdot)$ denotes global average pooling over time, and $D=128$ is the embedding dimension.
\vspace{-0.1cm}
\subsubsection{SSL-Feature-Based Gating}
\vspace{-0.1cm}
We use a pre-trained SSL model (XLSR~\cite{babu2021xls}) to extract frame-level representations $\bm{X}_{\text{ssl}} \in \mathbb{R}^{C \times L}$, where $C$ is the channel dimension and $L$ is the number of frames. Following RawNet2~\cite{rawnet2}, we apply a learnable 1D convolutional layer for channel mixing to obtain $\bm{X}'_{\text{ssl}} = \mathrm{Conv1D}(\bm{X}_{\text{ssl}}) \in \mathbb{R}^{D \times L}$.

We then apply the proposed depthwise Sinc layer (DW-Sinc Layer) to filter each channel of $\bm{X}'_{\text{ssl}}$ independently. DW-Sinc constructs a channel-wise Sinc filter bank $\{ \bm{h}_k(f_{1,k}, f_{2,k}) \}_{k=1}^{D}$, where each kernel is parameterized by channel-specific cutoff frequencies $(f_{1,k},f_{2,k})$. The depthwise convolution is computed as
\begin{equation}
  \tilde{\bm{X}}_{\text{ssl}}^{(k)}
  =
  \bm{X}'_{\text{ssl}}{}^{(k)} \circledast
  \bm{h}_k(f_{1,k}, f_{2,k}),
  \quad k = 1,\dots,D,
\end{equation}
where $\bm{X}'_{\text{ssl}}{}^{(k)} \in \mathbb{R}^{L}$ denotes the $k$-th channel of $\bm{X}'_{\text{ssl}}$ and $\tilde{\bm{X}}_{\text{ssl}}^{(k)} \in \mathbb{R}^{L}$ denotes the corresponding filtered sequence. Stacking all $\bm{X}'_{\text{ssl}}{}^{(k)}$ along the channel dimension yields $\tilde{\bm{X}}_{\text{ssl}} \in \mathbb{R}^{D \times L}$.

Finally, two stacked ResNet blocks model temporal context, and attentive pooling \cite{dos2016attentive} produces the SSL gating embedding:
\begin{equation}
  \bm{e}_{\text{ssl}}
  =
  \mathrm{AttentivePool}\left(
    \mathrm{ResNet}(\tilde{\bm{X}}_{\text{ssl}})
  \right) \in \mathbb{R}^{D},
\end{equation}
where $\mathrm{ResNet}(\cdot)$ denotes the stacked residual blocks and $\mathrm{AttentivePool}(\cdot)$ denotes attentive pooling over time.

\subsubsection{Prototype-Guided Routing}
This stage maps the gating embeddings $\bm{e}_{\text{raw}}$ and $\bm{e}_{\text{ssl}}$ to routing weights over $N$ experts. We compute two sets of expert logits: (i) content logits derived from the concatenated embeddings and (ii) prototype logits derived from the similarity between $\bm{e}_{\text{ssl}}$ and the learnable domain prototypes. The two logit vectors are then fused to produce the final gating weights.

We obtain content logits $\bm{l}_{\text{content}} \in \mathbb{R}^{N}$ by concatenating the two gating embeddings followed by a linear projection:
\begin{equation}
  \bm{l}_{\text{content}} =
  \text{Linear}\left(
    \text{Concat}(\bm{e}_{\text{raw}}, \bm{e}_{\text{ssl}})
  \right)
  \in \mathbb{R}^{N},
\end{equation}
where $\text{Concat}(\cdot,\cdot)$ denotes concatenation and $\text{Linear}(\cdot)$ denotes a learned linear projection.

To incorporate domain priors, we introduce a set of prototypes
$\{\bm{P}^{(i)}\}_{i=1}^{N}$, where $\bm{P}^{(i)} \in \mathbb{R}^{D}$.
Prototype logits $\bm{l}_{\text{proto}} \in \mathbb{R}^{N}$ are computed using cosine similarity between $\bm{e}_{\text{ssl}}$ and each prototype:
\begin{equation}
  l_{\text{proto}}^{(i)}
  =
  \cos(\bm{e}_{\text{ssl}}, \bm{P}^{(i)})
  \in \mathbb{R},
  \quad i \in \{1, \dots, N\},
\end{equation}
where $l_{\text{proto}}^{(i)}$ is the $i$-th logit of $\bm{l}_{\text{proto}}$.

The final gating logits $\bm{g} \in \mathbb{R}^{N}$ are obtained by a convex fusion of $\bm{l}_{\text{content}}$ and $\bm{l}_{\text{proto}}$:
\begin{equation}
  \bm{g}
  =
  \sigma(\bm{\alpha}_{g}) \cdot \bm{l}_{\text{content}}
  +
  \bigl(1 - \sigma(\bm{\alpha}_{g})\bigr)
  \cdot \bm{l}_{\text{proto}} \cdot \tau ,
  \label{eq:gating_logits_fusion}
\end{equation}
where $\sigma(\cdot)$ denotes the sigmoid function, $\bm{\alpha}_g$ is a learnable scalar, and $\tau$ is a temperature scalar. Routing weights $\bm{w} \in \mathbb{R}^{N}$ are then obtained via softmax normalization:
\begin{equation}
  \bm{w}
  =
  \mathrm{Softmax}(\bm{g})
  \in \mathbb{R}^{N}.
  \label{eq:routing_weights}
\end{equation}

\subsection{Affine Experts and Top-k Inference}
DADGMoE uses $N$ affine experts $\mathcal{E}=\{E_1, E_2, \dots, E_N \}$, where each expert $E_i$ is implemented as a BN layer with expert-specific affine parameters $(\bm{\gamma}_i,\bm{\beta}_i)$ and running statistics $(\bm{\mu}_i,\bm{\sigma}_i)$. Given the input SSL feature sequence $\bm{X}_{\text{ssl}} \in \mathbb{R}^{C \times L}$, expert $E_i$ applies the following transformation:
\begin{equation}
  E_i(\bm{X}_{\text{ssl}}) = \bm{\gamma}_i \odot \frac{\bm{X}_{\text{ssl}} - \bm{\mu}_i}{\sqrt{\bm{\sigma}_i^2 + \epsilon}} + \bm{\beta}_i ,
\end{equation}
where $\bm{\gamma}_i$ and $\bm{\beta}_i$ are the learnable scale and bias parameters, $(\bm{\mu}_i,\bm{\sigma}_i)$ are the running mean and standard deviation, and $\epsilon$ is a small constant for numerical stability.

To reduce computational cost and encourage expert specialization, we adopt a top-$k$ routing strategy that selects only the $k$ experts with the largest routing weights. The final output $\bm{Y} \in \mathbb{R}^{C \times L}$ of the DADGMoE module is the weighted sum of the selected experts' outputs:
\begin{equation}
  \bm{Y} = \sum_{i \in \text{top-}k} w_i E_i(\bm{X}_{\text{ssl}}),
\end{equation}
where $\bm{w}_i$ represents the $i$-th entry of $\bm{w}$ in Eq.~\ref{eq:routing_weights}, corresponding to the weight for expert $E_i$, $\text{top-}k$ denotes the index set of the $k$ largest entries of $\bm{w}$. The refined feature sequence $\bm{Y}$ is fed into the backend classifier (e.g., AASIST) to obtain the final prediction.

\section{Experiments and Analysis}
\subsection{Experimental Settings}
\vspace{-0.2cm}
\myparagraph{Datasets and Metric}
We utilized the ASVspoof 2019 LA (19LA) training set~\cite{asvspoof2019} for training all systems. This dataset comprises bonafide speech and spoofed samples generated by six diverse voice conversion and text-to-speech algorithms. To rigorously evaluate domain generalization capabilities, we tested our models on three distinct evaluation sets representing various domain shifts: 21DF~\cite{yamagishi2021asvspoof}, ITW~\cite{muller2022itw}, FoR~\cite{reimao2019for}, and ADD 2023 test sets (R1 \& R2) (ADDR1 \& ADDR2)~\cite{yi2023add}.
Performance is reported using the equal error rate EER.

\myparagraph{Models}
Our experiments cover both standard supervised architectures and pre-trained SSL models. We use XLSR~\cite{babu2021xls}, a multilingual model optimized for cross-lingual tasks. These models serve as feature extractors and are fine-tuned with either the AASIST backend (XLSR-AASIST)~\cite{tak2022xlsraasist}
for classification.

\myparagraph{Implementation Details}
During training, audio segments were cropped or concatenated to approximately 4 seconds (64,600 samples). We optimized the models using a cross-entropy loss and the Adam optimizer \cite{kingma2014adam}. The learning rate was set to $1e^{-6}$ with a weight decay of $1e^{-5}$.
Rawboost is used for data augmentation~\cite{rawboost}.
A batch size of 20 is employed for all training steps. To ensure robust performance and prevent overfitting, early stopping is applied if the validation set's cross-entropy loss does not improve for 7 consecutive epochs.
Model checkpoints are generated by averaging the top-5 best performing models on the validation set, from which the final results are reported.
All experiments are conducted on a single Nvidia 4090 GPU, ensuring reproducibility by utilizing a consistent random seed across all runs.
\begin{table}[!t]
  \centering
  \caption{Overall SDD performance (EER\%) and parameter efficiency of the proposed DADGMoE framework compared to the baseline Model. * denotes results reported by~\cite{dowerah2026board}.}
  \label{tab:overall_performance}
  \resizebox{0.95\linewidth}{!}{
    \begin{tabular}{lcccccc}
      \toprule
      \textbf{Model} & \textbf{Params (M)} & \textbf{21DF} & \textbf{ITW} & \textbf{FoR} & \textbf{ADDR1} & \textbf{ADDR2}\\
      \midrule
      XLSR-AASIST & 317.84 & 3.69 & 10.46 & 7.47* & 27.74* & 21.93* \\
      \textbf{DADGMoE} & 318.01 & \textbf{2.54} & \textbf{6.35} & \textbf{4.42} & \textbf{23.85}& \textbf{21.61}\\
      \bottomrule
  \end{tabular}}
\end{table}

\begin{table}[t]
  \centering
  \caption{Ablation Study of DADGMoE Components (EER\%). The full model is DADGMoE (5 experts; top-k=2). Values in parentheses indicate the absolute EER difference compared to the proposed method. * denotes results reported by~\cite{dowerah2026board}}
  \label{tab:ablation_study_plus_minus}
  \resizebox{0.90\linewidth}{!}{
    \begin{tabular}{lccc}
      \toprule
      \textbf{Model} & \textbf{21DF} & \textbf{ITW} & \textbf{FoR} \\
      \midrule
      XLSR-AASIST & 3.69 & 10.46 & 7.47*  \\
      \midrule
      \textbf{DADGMoE } & \textbf{2.54} & \textbf{6.35} & \textbf{4.42} \\
      \quad w/o Prototypes & 2.54 \textbf{(+0.00)} & 8.46 \textbf{(+2.11)} & 5.31 \textbf{(+0.89)} \\
      \quad w/o SSL-Gating & 2.66 \textbf{(+0.12)} & 8.06 \textbf{(+1.71)} & 6.67 \textbf{(+2.25)} \\
      \quad w/o Raw-Gating & 3.89 \textbf{(+1.35)} & 9.88 \textbf{(+3.53)} & 12.67 \textbf{(+8.25)} \\
      \bottomrule
  \end{tabular}}
\end{table}

\begin{table}[t]
  \centering
  \caption{Impact of Top-$k$ Routing Strategy for DADGMoE (5 experts) on Performance (EER\%).}
  \label{tab:topk_analysis}
  \resizebox{0.5\linewidth}{!}{
    \begin{tabular}{lccc}
      \toprule
      \textbf{Top-$k$ Value} & \textbf{21DF} & \textbf{ITW} & \textbf{FoR} \\
      \midrule
      $k=1$ & 2.54 & 8.35 & 6.24 \\
      $k=2$ & \textbf{2.54} & \textbf{6.35} & \textbf{4.42} \\
      $k=3$ & 3.19 & 8.70 & 9.32 \\
      $k=4$ & 2.40 & 6.85 & 5.44 \\
      $k=5$ & 2.57 & 7.20 & 6.45 \\
      \bottomrule
  \end{tabular}}
\end{table}

\begin{table}[htbp]
  \centering
  \caption{Comparison of DADGMoE (5 experts; top-k = 2) with state-of-the-art SDD Systems (EER\%). * denotes results reported in~\cite{dowerah2026board}.} 
  \label{tab:sota_comparison_styled}
  \resizebox{0.88\linewidth}{!}{
    \begin{tabular}{lccc}
      \toprule
      \textbf{System} & \textbf{21DF} & \textbf{ITW} & \textbf{FoR} \\
      \midrule
      XLSR-MoE~\cite{moe2025icassp} & 2.54 & 9.17 & - \\
      XLSR-SLS~\cite{zhang2024sls} & 1.92 & 7.46 & 5.07* \\
      XLSR-Nes2Net-X~\cite{liu2025nes2net} & 1.78 & 6.60 & 6.31* \\
      XLSR-AASIST-SAM~\cite{huang25sharp} & 3.44 & 6.34 & 5.18 \\
      Wav2DF-TSL~\cite{hao2025moeijcnn} & 1.95 & 6.83 & - \\
      XLSR-Conformer-TCM~\cite{truong24tcm} & 2.06 & 7.79 & 10.68* \\
      \midrule
      XLSR-AASIST~\cite{tak2022xlsraasist} & 3.69 & 10.46 & 7.47* \\
      \quad + DADGMoE & \textbf{2.54} & \textbf{6.35} & \textbf{4.42} \\ 
      \bottomrule
  \end{tabular}}
\end{table}

\vspace{-0.1cm}
\subsection{Overall Performance and Efficiency}
\vspace{-0.1cm}
We first evaluate the effectiveness of our proposed DADGMoE framework by comparing it against the strong baseline, XLSR-AASIST. As shown in Table~\ref{tab:overall_performance}, our method achieves substantial performance gains across all three benchmarks.
On the standard 21DF dataset, DADGMoE reduces the EER from 3.69\% to \textbf{2.54\%}, representing a relative improvement of \textbf{31.2\%}. More importantly, on the challenging out-of-domain datasets, the improvements are even more significant. On the ITW dataset, the EER drops significantly from 10.46\% to \textbf{6.35\%} (a \textbf{39.3\%} reduction). Similarly, on the FoR dataset, we observe a \textbf{40.8\%} relative reduction in EER (7.47\% to \textbf{4.42\%}). On the two ADD 2023 test sets, we also observe a reduction in EER.

Crucially,
the DADGMoE introduces only \textbf{0.17M} additional parameters, as highlighted in the parameter comparison in Table~\ref{tab:overall_performance}. This validates the efficiency of our lightweight affine expert design.
\vspace{-0.1cm}
\subsection{Ablation Studies}
\vspace{-0.2cm}
\myparagraph{Impact of Dual-Gating Mechanism}
In Table~\ref{tab:ablation_study_plus_minus}, comparing the single-branch variants reveals the necessity of our gating strategy. Removing the Raw-Gating branch ( relying only on SSL features, denoted as ``w/o Raw-Gating'') leads to a performance drop, particularly on the FoR dataset, where the EER increases by \textbf{8.25\%}. This confirms that high-level SSL features alone may miss critical low-level artifacts. Conversely, removing the SSL-Gating branch (``w/o SSL-Gating'') also degrades performance (e.g., +1.71\% on ITW), indicating that raw waveforms lack the semantic context required for robust detection. The full dual-gating model effectively synergizes both levels.

\myparagraph{Effectiveness of Domain Prototypes}
Removing the domain prototypes (``w/o Prototypes'') results in a consistent performance degradation, most notably on the cross-domain ITW dataset (+2.11\% EER), as detailed in Table~\ref{tab:ablation_study_plus_minus}. This empirical evidence supports our hypothesis that learnable prototypes act as essential semantic anchors, guiding the router to generalize better to unseen domains by matching implicit deepfake patterns.
\vspace{-0.3cm}
\subsection{Analysis of Routing Strategy and SOTA Comparison}
\vspace{-0.2cm}
\myparagraph{Top-$k$ Analysis}
We further investigate the impact of the expert activation count $k$. As shown in Table~\ref{tab:topk_analysis}, setting $k=2$ yields the optimal trade-off between specialization and diversity. A single expert ($k=1$) may lack sufficient capacity to handle complex attacks, while activating too many experts ($k \ge 3$) appears to reduce domain-specific specialization, and adding more affine normalizers may cause scaling conflicts.

\myparagraph{Comparison with State-of-the-Art}
Table~\ref{tab:sota_comparison_styled} compares DADGMoE with recent state-of-the-art systems. Our method outperforms several competitive MoE and SSL-based approaches, including XLSR-MoE~\cite{moe2025icassp} and Wav2DF-TSL~\cite{hao2025moeijcnn}, particularly on the ITW and FoR benchmark.
\begin{figure}[t]
  \centering
  \includegraphics[width=0.7\columnwidth]{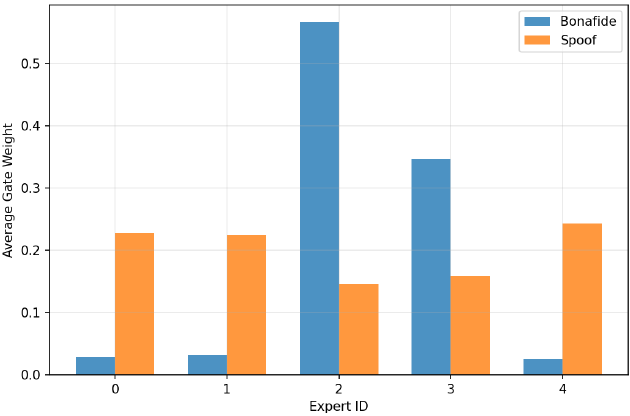} 
  \caption{Expert specialization in DADGMoE. Average gate weights for bonafide and spoof samples from the training set across experts, demonstrating clear roles (e.g., Expert 2 for bonafide, Expert 4 for spoof).}
  \label{fig:expert_vis}
\end{figure}

\begin{figure}[t]
  \centering
  \includegraphics[width=0.7\columnwidth]{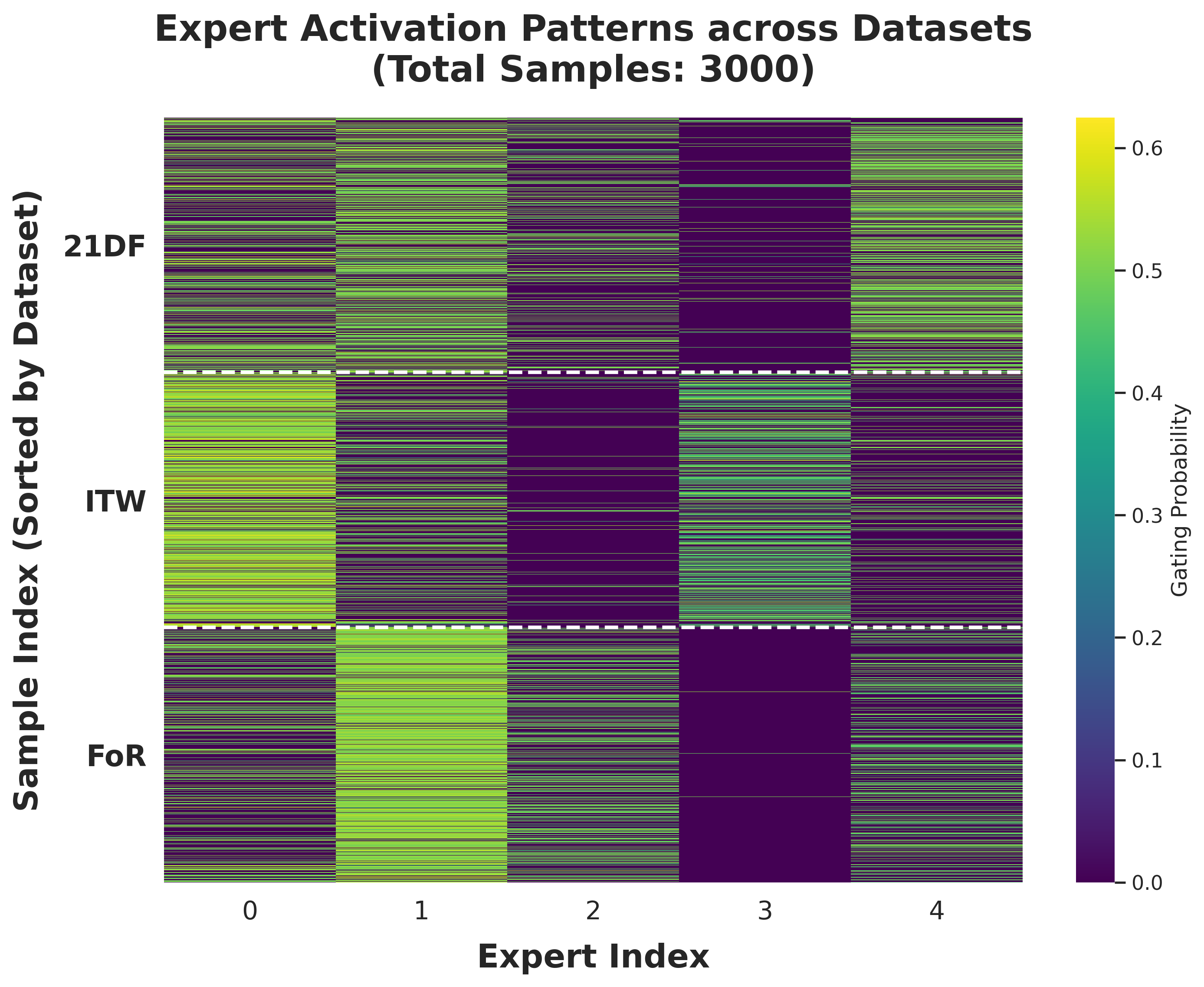} 
  \caption{The heatmap visualizes the gating probability for each expert across samples from 21DF, ITW, and FoR datasets. Clear vertical banding patterns indicate expert specialization towards specific deepfake domains.}
  \label{fig:expert_domain_specialization}
\end{figure}
\vspace{-0.1cm}
\subsection{Analysis of Expert Specialization and Domain Adaptation}
\vspace{-0.2cm}
Figure~\ref{fig:expert_vis} visualizes the average gate weight distribution for bonafide and spoof samples across experts, revealing clear class-based specialization. Expert 2 predominantly activates for bonafide samples, while Experts 0, 1, and 4 exhibit higher engagement for spoofed samples, likely covering distinct deepfake sub-patterns. This provides qualitative evidence that our lightweight affine experts effectively adapt to and capture diverse deepfake characteristics.

Further, Figure~\ref{fig:expert_domain_specialization} presents a heatmap of expert activation patterns across different datasets, demonstrating compelling evidence for domain-specific expert specialization. Expert 1 and 4 primarily activate for 21DF samples, while Expert 0 shows predominant activation for ITW samples. This specialized routing strategy, informed by our learned domain prototypes, confirms that our gating mechanism effectively directs inputs from different deepfake domains to different experts.
\vspace{-0.2cm}
\section{Conclusion}
\vspace{-0.1cm}
We here introduced the DADGMoE framework, a novel approach designed to enhance the generalization capabilities of audio deepfake detection systems.
Through extensive experiments on challenging ASVspoof 2021 benchmarks, DADGMoE achieved significant performance gains, notably reducing the EER by up to 40.8\% on out-of-domain datasets.

\section{Acknowledgment}
This work was supported in part by the Innovation and Technology Fund of the Hong Kong SAR (Project No. MHP/048/24), and the National Key R\&D Program of China (2024YFE0217200).

\section{Use of Generative AI Disclosure}
\label{GenAI}
Generative AI tools were used only for language polishing and formatting assistance. All scientific content, experiments, and analyses were produced and verified by the authors.

\bibliographystyle{IEEEtran}
\bibliography{mybib_moe}

@article{kingma2014adam,
  title={Adam: A method for stochastic optimization},
  author={Kingma, Diederik P and Ba, Jimmy},
  journal={arXiv preprint arXiv:1412.6980},
  year={2014}
}

@article{dos2016attentive,
  title={Attentive pooling networks},
  author={Dos Santos, C{\i}cero Nogueira and Tan, Ming and Xiang, Bing and Zhou, Bowen},
  journal={arXiv preprint arXiv:1602.03609},
  volume={8},
  year={2016}
}

@inproceedings{wang2025moe,
  title={Mixture of experts fusion for fake audio detection using frozen wav2vec 2.0},
  author={Wang, Zhiyong and Fu, Ruibo and Wen, Zhengqi and Tao, Jianhua and Wang, Xiaopeng and Xie, Yuankun and Qi, Xin and Shi, Shuchen and Lu, Yi and Liu, Yukun and others},
  booktitle={Proc. ICASSP},
  pages={1--5},
  year={2025},
}

@inproceedings{chang2019bn,
  title={Domain-specific batch normalization for unsupervised domain adaptation},
  author={Chang, Woong-Gi and You, Tackgeun and Seo, Seonguk and Kwak, Suha and Han, Bohyung},
  booktitle={Proc. IEEE/CVF conference on Computer Vision and Pattern Recognition},
  pages={7354--7362},
  year={2019}
}

@inproceedings{wu2024testbn,
  title={Test-time domain adaptation by learning domain-aware batch normalization},
  author={Wu, Yanan and Chi, Zhixiang and Wang, Yang and Plataniotis, Konstantinos N and Feng, Songhe},
  booktitle={Proceedings of the AAAI Conference on Artificial Intelligence},
  volume={38},
  number={14},
  pages={15961--15969},
  year={2024}
}

@INPROCEEDINGS{sincnet2,
  author={Ho, Kuan-Hsun and Hung, Jeih-weih and Chen, Berlin},
  booktitle={Proc. ICASSP}, 
  title={What Do Neural Networks Listen to? Exploring the Crucial Bands in Speech Enhancement Using SINC-Convolution}, 
  year={2024},
  volume={},
  number={},
  pages={10406-10410},
  doi={10.1109/ICASSP48485.2024.10445878}}

@INPROCEEDINGS{sincnet,
  author={Ravanelli, Mirco and Bengio, Yoshua},
  booktitle={Proc. SLT}, 
  title={Speaker Recognition from Raw Waveform with SincNet}, 
  year={2018},
  volume={},
  number={},
  pages={1021-1028},
  doi={10.1109/SLT.2018.8639585}}

@article{moe2025survey,
  title={A survey on mixture of experts in large language models},
  author={Cai, Weilin and Jiang, Juyong and Wang, Fan and Tang, Jing and Kim, Sunghun and Huang, Jiayi},
  journal={IEEE Transactions on Knowledge and Data Engineering},
  year={2025},
  publisher={IEEE}
}

@inproceedings{hao2025moeijcnn,
  title={Wav2df-tsl: Two-stage learning with efficient pre-training and hierarchical experts fusion for robust audio deepfake detection},
  author={Hao, Yunqi and Chen, Yihao and Xu, Minqiang and Zhan, Jianbo and He, Liang and Fang, Lei and Fang, Sian and Liu, Lin},
  booktitle={Proc. IJCNN},
  pages={1--8},
  year={2025},
  organization={IEEE}
}

@inproceedings{laakkonen2025moe,
  title={Mixture of Low-Rank Adapter Experts in Generalizable Audio Deepfake Detection},
  author={Laakkonen, Janne and Kukanov, Ivan and Hautam{\"a}ki, Ville},
  booktitle={Proc. APSIPA ASC 2025},
  pages={2211--2216},
  year={2025},
  organization={IEEE}
}

@article{pan2025moe,
  title={MoLEx: Mixture of LoRA Experts in Speech Self-Supervised Models for Audio Deepfake Detection},
  author={Pan, Zihan and Bhupendra, Sailor Hardik and Wu, Jinyang},
  journal={arXiv preprint arXiv:2509.09175},
  year={2025}
}

@INPROCEEDINGS{moe2025icassp,
  author={Negroni, Viola and Salvi, Davide and Mezza, Alessandro Ilic and Bestagini, Paolo and Tubaro, Stefano},
  booktitle={ICASSP 2025 - 2025 IEEE International Conference on Acoustics, Speech and Signal Processing (ICASSP)}, 
  title={Leveraging Mixture of Experts for Improved Speech Deepfake Detection}, 
  year={2025},
  volume={},
  number={},
  pages={1-5},
  doi={10.1109/ICASSP49660.2025.10890398}}

@inproceedings{huang25sharp,
  title     = {{From Sharpness to Better Generalization for Speech Deepfake Detection}},
  author    = {Wen Huang and Xuechen Liu and Xin Wang and Junichi Yamagishi and Yanmin Qian},
  year      = {2025},
  booktitle = {{Proc. Interspeech 2025}},
  pages     = {5338--5342},
  doi       = {10.21437/Interspeech.2025-1095},
  issn      = {2958-1796},
}

@article{truong2025addressing,
  title={Addressing Gradient Misalignment in Data-Augmented Training for Robust Speech Deepfake Detection},
  author={Truong, Duc-Tuan and Liu, Tianchi and Li, Junjie and Tao, Ruijie and Lee, Kong Aik and Chng, Eng Siong},
  journal={arXiv preprint arXiv:2509.20682},
  year={2025}
}

@inproceedings{zhang2024sls,
  title={Audio deepfake detection with self-supervised XLS-R and SLS classifier},
  author={Zhang, Qishan and Wen, Shuangbing and Hu, Tao},
  booktitle={Proc. 32nd ACM International Conference on Multimedia},
  pages={6765--6773},
  year={2024}
}

@article{liu2025nes2net,
  title={Nes2net: A lightweight nested architecture for foundation model driven speech anti-spoofing},
  author={Liu, Tianchi and Truong, Duc-Tuan and Das, Rohan Kumar and Lee, Kong Aik and Li, Haizhou},
  journal={arXiv preprint arXiv:2504.05657},
  year={2025}
}

@article{dowerah2026board,
  title={Speech df arena: A leaderboard for speech deepfake detection models},
  author={Dowerah, Sandipana and Kulkarni, Atharva and Kulkarni, Ajinkya and Tran, Hoan My and Kalda, Joonas and Fedorchenko, Artem and Fauve, Benoit and Lolive, Damien and Alum{\"a}e, Tanel and Doss, Mathew Magimai-},
  journal={IEEE Open Journal of Signal Processing},
  year={2026},
  publisher={IEEE}
}

@article{tak2022xlsraasist,
  title={Automatic speaker verification spoofing and deepfake detection using wav2vec 2.0 and data augmentation},
  author={Tak, Hemlata and Todisco, Massimiliano and Wang, Xin and Jung, Jee-weon and Yamagishi, Junichi and Evans, Nicholas},
  journal={arXiv preprint arXiv:2202.12233},
  year={2022}
}

@inproceedings{truong24tcm,
  title     = {{Temporal-Channel Modeling in Multi-head Self-Attention for Synthetic Speech Detection}},
  author    = {Duc-Tuan Truong and Ruijie Tao and Tuan Nguyen and Hieu-Thi Luong and Kong Aik Lee and Eng Siong Chng},
  year      = {2024},
  booktitle = {{Proc. Interspeech 2024}},
  pages     = {537--541},
  doi       = {10.21437/Interspeech.2024-659},
  issn      = {2958-1796},
}

@inproceedings{reimao2019for,
  title={For: A dataset for synthetic speech detection},
  author={Reimao, Ricardo and Tzerpos, Vassilios},
  booktitle={Proc.  Speech Technology and Human-Computer Dialogue (SpeD)},
  pages={1--10},
  year={2019},
  organization={IEEE}
}

@article{muller2022itw,
  title={Does audio deepfake detection generalize?},
  author={M{\"u}ller, Nicolas M and Czempin, Pavel and Dieckmann, Franziska and Froghyar, Adam and B{\"o}ttinger, Konstantin},
  journal={arXiv preprint arXiv:2203.16263},
  year={2022}
}

@article{yi2023add,
  title={Add 2023: the second audio deepfake detection challenge},
  author={Yi, Jiangyan and Tao, Jianhua and Fu, Ruibo and Yan, Xinrui and Wang, Chenglong and Wang, Tao and Zhang, Chu Yuan and Zhang, Xiaohui and Zhao, Yan and Ren, Yong and others},
  journal={arXiv preprint arXiv:2305.13774},
  year={2023}
}

@article{babu2021xls,
  title={XLS-R: Self-supervised cross-lingual speech representation learning at scale},
  author={Babu, Arun and Wang, Changhan and Tjandra, Andros and Lakhotia, Kushal and Xu, Qiantong and Goyal, Naman and Singh, Kritika and Von Platen, Patrick and Saraf, Yatharth and Pino, Juan and others},
  journal={arXiv preprint arXiv:2111.09296},
  year={2021}
}

@ARTICLE{asvspoof2019,
  author={Nautsch, Andreas and Wang, Xin and Evans, Nicholas and Kinnunen, Tomi H. and Vestman, Ville and Todisco, Massimiliano and Delgado, Héctor and Sahidullah, Md and Yamagishi, Junichi and Lee, Kong Aik},
  journal={IEEE Transactions on Biometrics, Behavior, and Identity Science}, 
  title={{ASV}spoof 2019: Spoofing Countermeasures for the Detection of Synthesized, Converted and Replayed Speech}, 
  year={2021},
  volume={3},
  number={2},
  pages={252-265},
  doi={10.1109/TBIOM.2021.3059479}}

@inproceedings{yamagishi2021asvspoof,
  title={{ASV}spoof 2021: Accelerating progress in spoofed and deepfake speech detection},
  author={Yamagishi, Junichi and Wang, Xin and Todisco, Massimiliano and Sahidullah, Md and Patino, Jose and Nautsch, Andreas and Liu, Xuechen and Lee, Kong Aik and Kinnunen, Tomi and Evans, Nicholas and others},
  booktitle={ASVspoof 2021 Workshop},
  year={2021}
}

@inproceedings{asvspoof5,
    author = {Wang, Xin and Delgado, H{\'e}ctor and Tak, Hemlata and Jung, Jee-weon and Shim, Hye-jin and Todisco, Massimiliano and Kukanov, Ivan and Liu, Xuechen and Sahidullah, Md and Kinnunen, Tomi and others},
    title = {{ASV}spoof 5: Crowdsourced speech data, deepfakes, and adversarial attacks at scale},
    booktitle = {Proc. INTERSPEECH 2024},
    year = {2024},
    pages={1-8}
}

@INPROCEEDINGS{add2022,
  author={Yi, Jiangyan and Fu, Ruibo and Tao, Jianhua and Nie, Shuai and Ma, Haoxin and Wang, Chenglong and Wang, Tao and Tian, Zhengkun and Bai, Ye and Fan, Cunhang and Liang, Shan and Wang, Shiming and Zhang, Shuai and Yan, Xinrui and Xu, Le and Wen, Zhengqi and Li, Haizhou},
  booktitle={Proc. IEEE International Conference on Acoustics, Speech and Signal Processing (ICASSP)}, 
  title={{ADD} 2022: The first Audio Deep Synthesis Detection Challenge}, 
  year={2022},
  volume={},
  number={},
  pages={9216-9220},
  doi={10.1109/ICASSP43922.2022.9746939}}

@INPROCEEDINGS{rawnet2,
  author={Tak, Hemlata and Patino, Jose and Todisco, Massimiliano and Nautsch, Andreas and Evans, Nicholas and Larcher, Anthony},
  booktitle={Proc. IEEE International Conference on Acoustics, Speech and Signal Processing (ICASSP)}, 
  title={End-to-End anti-spoofing with {RawNet2}}, 
  year={2021},
  volume={},
  number={},
  pages={6369-6373},
  doi={10.1109/ICASSP39728.2021.9414234}}

@INPROCEEDINGS{rawboost,
  author={Tak, Hemlata and Kamble, Madhu and Patino, Jose and Todisco, Massimiliano and Evans, Nicholas},
  booktitle={Proc. ICASSP 2022}, 
  title={Rawboost: A Raw Data Boosting and Augmentation Method Applied to Automatic Speaker Verification Anti-Spoofing}, 
  year={2022},
  volume={},
  number={},
  pages={6382-6386},
  doi={10.1109/ICASSP43922.2022.9746213}}

@article{aasist2024,
  title={{AASIST3}: KAN-Enhanced {AASIST} Speech Deepfake Detection using {SSL} Features and Additional Regularization for the {ASV}spoof 2024 Challenge},
  author={Borodin, Kirill and Kudryavtsev, Vasiliy and Korzh, Dmitrii and Efimenko, Alexey and Mkrtchian, Grach and Gorodnichev, Mikhail and Rogov, Oleg Y},
  journal={arXiv preprint arXiv:2408.17352},
  year={2024}
}

@ARTICLE{crs_ds3,
  author={Liu, Yuchen and Chen, Yabo and Dai, Wenrui and Gou, Mengran and Huang, Chun-Ting and Xiong, Hongkai},
  journal={IEEE Transactions on Pattern Analysis and Machine Intelligence}, 
  title={Source-Free Domain Adaptation With Domain Generalized Pretraining for Face Anti-Spoofing}, 
  year={2024},
  volume={46},
  number={8},
  pages={5430-5448},
  doi={10.1109/TPAMI.2024.3370721}}

@article{li2026towards,
  author={Li, Zhe and Mak, Man-Wai and Pilanci, Mert and Lee, Hung-Yi and Gan, Chong-Xin and Sheng, Jiabao and Meng, Helen},
  title={Towards a Unified Perspective on Parameter-Efficient Fine-Tuning for Speaker Verification},
  journal={IEEE Transactions on Audio, Speech, and Language Processing},
  volume={34},
  pages={2276--2289},
  year={2026}
}

\end{document}